\documentstyle[twocolumn,eqsecnum,epsfig,aps]{revtex}

\begin{document}
\draft

\title{Collective effects in the collapse-revival phenomenon and
squeezing in the Dicke model}
\author{G. Ramon, \cite{email1} C. Brif, \cite{email2} 
and A. Mann \cite{email3}}
\address{Department of Physics, Technion -- Israel Institute of 
Technology, Haifa 32000, Israel}
   \maketitle

        \begin{abstract}
Resonant interaction of a collection of two-level atoms with a 
single-mode coherent cavity field is considered in the framework 
of the Dicke model. We focus on the role of collective atomic
effects in the phenomenon of collapses and revivals of the Rabi
oscillations. It is shown that the behavior of the system strongly
depends on the initial atomic state. In the case of the initial
half-excited Dicke state we account for a number of interesting 
phenomena. The correlations between the atoms result 
in a suppression of the revival amplitude, and the revival time 
is halved, compared to the uncorrelated fully-excited and ground 
states. The phenomenon of squeezing of the radiation field in the 
atom-field interaction is also discussed. For the initial 
fully-excited and ground atomic states, the field is squeezed on 
the short-time scale, and squeezing can be enhanced by increasing 
the number of atoms. Some empirical formulas are found which 
describe the behavior of the system in excellent agreement with 
numerical results. For the half-excited Dicke state, the field 
can be strongly squeezed on the long-time scale in the case of 
two atoms. This kind of squeezing is enhanced by increasing the 
intensity of the initial coherent field and is of the same nature 
as revival-time squeezing in the Jaynes-Cummings model. 
The appearance of this long-time squeezing can be explained using 
the factorization approximation for semiclassical atomic states.
        \end{abstract}


\section{Introduction}
\label{sec:intr}

Since the pioneering work of Dicke \cite{Dicke54} on cooperative
spontaneous emission, a great deal of attention has been devoted 
to the interaction of the radiation field with a collection of 
two-level atoms located within a distance much smaller than the 
wavelength of the radiation. Such a system is commonly referred to 
as the Dicke model (for a review see, e.g., Ref.\ \cite{LePeVe86}). 
A particular case of the Dicke model, when atoms interact with a 
single-mode radiation field inside a cavity, was considered by 
Tavis and Cummings \cite{TC68}. The Tavis-Cummings Hamiltonian 
is mathematically equivalent to the trilinear boson Hamiltonian 
describing various nonlinear optical processes \cite{WoBa70}. 
The single-atom version, known as the Jaynes-Cummings model 
\cite{JC63}, is the simplest and one of the most popular models of 
quantum optics. In spite of its simplicity, the Jaynes-Cummings 
model shows a variety of interesting nonclassical phenomena such as 
vacuum-field Rabi oscillations, sub-Poissonian photon statistics, 
and squeezing of the radiation field (for reviews see, e.g., 
Refs.\ \cite{KnWe92,ShKn93}).

One of the most interesting quantum features of the Jaynes-Cummings 
model is the phenomenon of collapses and revivals of the Rabi
oscillations, which manifests itself in the clearest way when the 
cavity field is prepared initially in the coherent state 
\cite{EbNaSM}. The shape of collapses and revivals is determined
by the initial photon-number distribution. A similar behavior can 
be found also in the many-atom case \cite{BaKn84,DroJex,ChKlSM94}. 
For a sufficiently strong coherent field, the nonlinearity in the 
Rabi frequency is slight and the system exhibits regular dynamics 
in the form of collapses and revivals of the oscillations. 
However, in the many-atom case there exist anharmonic collective 
corrections which modify the shape of the collapses and revivals 
related to the photon-distribution mechanism \cite{ChKo96}. 
In the present work we study in detail these collective effects 
for different initial atomic states. If atoms are prepared 
initially in the ground state or in the fully-excited state, then 
the system behaves rather similarly to the single-atom case, 
although collective effects manifest themselves clearly in the 
length of the revival time.
But, the results are different when the atoms are prepared 
initially in the half-excited Dicke state. This state is well 
known as the superradiant atomic state in the context of 
collective spontaneous emission in open space 
\cite{Dicke54,LePeVe86}. It was also found \cite{CBCSM96} that 
the half-excited Dicke state strongly modifies the resonance 
fluorescence spectrum of the atomic system.
However, to the best of our knowledge, 
this state was rarely considered in the context of the 
Tavis-Cummings (cavity) version of the Dicke model. 
When the atoms are prepared in the half-excited Dicke state 
the system exhibits a number of interesting phenomena. 
In particular, the amplitude of the Rabi oscillations is strongly 
suppressed and relative intensities of revivals are essentially 
changed. This suppresion of the revival amplitude is similar to 
the trapping phenomenon occuring in the case of a single atom 
prepared in the equally weighted superposition of the two 
levels \cite{ZaZu89}.

It has been known for a long time \cite{MeZu82,ShFLKAl87} that 
the nonlinear character of the Jaynes-Cummings model leads to 
squeezing in one of the quadratures of an initially coherent 
cavity field. It was also predicted \cite{KuMa88} that strong 
squeezing can be obtained in the Jaynes-Cummings model near the 
revival times for large initial intensities of the field.
This phenomenon of strong revival-time squeezing was explained 
in Ref.\ \cite{WoGB93} using the factorization approximation for 
the so-called semiclassical atomic states \cite{GB91}. 
Butler and Drummond \cite{BuDr86} showed that short-time squeezing 
can be enhanced in the Dicke model compared to the single-atom 
case. Seke \cite{Seke:sq} considered field and atomic squeezing in 
the Dicke model without the rotating-wave approximation and in the 
presence of losses.
Analytical approaches to squeezing in the strong-field limit 
\cite{RSKC97} and in the weak-field limit \cite{KoCh97} were
presented. Higher-order squeezing in the Dicke model was studied 
in Ref.\ \cite{Li90}. 
For atoms prepared initially in the ground state or in the 
fully-excited state, the uncertainty of the field quadrature 
rapidly oscillates and squeezing occurs during short periods 
of time at the very beginning of the time evolution. On the
long-time scale the temporal behavior of the quadrature 
uncertainty is correlated with times of collapses and revivals
of the Rabi oscillations.
On the other hand, for the initial half-excited Dicke state, the 
behavior of squeezing is qualitatively different. The quadrature 
uncertainty oscillates on the long-time scale, with a period of 
the order of the revival half-time, and strong squeezing is 
obtained in the case of two atoms. The value of squeezing is 
enhanced by the intensity of the initial coherent field.
This phenomenon is of the same nature as revival-time squeezing
in the Jaynes-Cummings model and can be explained using the 
factorization approximation for the semiclassical atomic states.
In general, squeezing is better and its duration is much longer 
for the initial two-atom half-excited Dicke state than for the 
ground and fully-excited states.

\section{The model and methods of solution}

We consider the resonant interaction between $N$ two-level atoms
and the single-mode radiation field inside a lossless cavity. 
In the rotating-wave approximation, the Tavis-Cummings 
interaction Hamiltonian reads ($\hbar = 1$):
\begin{equation}
H = g (a^{\dagger} J_{-} + a J_{+}) .   \label{Ham}
\end{equation}
Here and in the following we use the interaction picture.
In Eq.\ (\ref{Ham}) $g$ is the coupling constant, $a$ and 
$a^{\dagger}$ are the annihilation and creation operators of the 
field mode, $J_{+}$ and $J_{-}$ are the collective atomic raising 
and lowering operators. They satisfy the su(2) Lie algebra, 
\begin{equation}
[J_{+},J_{-}] = 2J_{z} , \;\;\;\;\;\;\; 
[J_{z},J_{\pm}] = \pm J_{\pm} ,
\end{equation}
where $J_{z}$ is the operator of atomic inversion. In terms of 
the standard Pauli matrices, describing each two-level atom, 
one obtains
\begin{equation}
J_{\pm} = \frac{1}{2} \sum_{i=1}^{N} \sigma^{(i)}_{\pm} ,
\;\;\;\;\;\;\;\;\;
J_{z} = \frac{1}{2} \sum_{i=1}^{N} \sigma^{(i)}_{z} .
\label{atreal} 
\end{equation}
If the Schwinger realization of the SU(2) generators is used,
the Hamiltonian (\ref{Ham}) becomes the trilinear boson 
Hamiltonian describing nonlinear optical processes such as
parametric conversion and Raman and Brillouin scattering 
\cite{WoBa70}. An interesting physical realization of the 
Hamiltonian (\ref{Ham}) is given by the coupling of the 
internal levels of atoms or ions to a mode of their quantized 
oscillatory motion in a harmonic trap \cite{WBIH94}.  

The total excitation operator
\begin{equation}
L = a^{\dagger} a + J_{z} + N/2  
\end{equation}
commutes with the Hamiltonian (\ref{Ham}) and is an integral
of motion. Another integral of motion is the SU(2) Casimir
operator 
${\bf J}^2 = J_{z}^2 + \frac{1}{2} (J_{+}J_{-} + J_{-}J_{+})$.
We describe the state of the atomic system in terms of the
SU(2) orthonormal basis $|j,m\rangle_{{\rm at}}$
($m=j,j-1,\ldots,-j$),
\begin{eqnarray}
& & J_{z} |j,m\rangle_{{\rm at}} = m |j,m\rangle_{{\rm at}} , \\
& & {\bf J}^2 |j,m\rangle_{{\rm at}} = j(j+1) 
|j,m\rangle_{{\rm at}} .
\end{eqnarray}
In the context of the atomic realization (\ref{atreal}), the
states $|j,m\rangle_{{\rm at}}$ are the symmetric Dicke states: 
\begin{equation}
|j,m\rangle_{{\rm at}} = \left( 
  \begin{array}{c}
  N \\ p
  \end{array}
\right)^{-1/2} \sum \prod_{k=1}^{p} |+\rangle_{l_{k}} 
\prod_{l \neq l_{k}} |-\rangle_{l} ,
\label{dstates}
\end{equation}
where $|+\rangle_{l}$ and $|-\rangle_{l}$ are the upper and 
lower states, respectively, of the $l$th atom, and the summation 
is over all possible permutations of $N$ atoms. 
If only symmetric atomic states are considered, then the 
`cooperative number' $j$ is equal to $N/2$ and $p=m+j$ is just
the number of excited atoms.

The Hilbert space ${\cal H}$ of the atom-field system can be 
decomposed into a direct sum of finite-dimensional invariant
subspaces ${\cal H}_{L}$:
\begin{equation}
{\cal H} = \bigoplus_{L=0}^{\infty} {\cal H}_{L} .
\end{equation}
Each invariant subspace ${\cal H}_{L}$ is spanned by the 
orthonormal basis $|n\rangle_{{\rm f}} 
|j,L-j-n\rangle_{{\rm at}}$, 
where $|n\rangle_{{\rm f}}$ are the Fock states of the
radiation field, $a^{\dagger} a |n\rangle_{{\rm f}} =
n |n\rangle_{{\rm f}}$. For $L<N$, $n=0,1,\ldots,L$ and
${\rm dim}({\cal H}_{L}) = L+1$; for $L \geq N$, 
$n=L-N,L-N+1,\ldots,L$ and ${\rm dim}({\cal H}_{L}) = N+1$.
If the field is initially in the Fock state 
$|n_{0}\rangle_{{\rm f}}$ and the atoms are in the Dicke
state $|j,m_{0}\rangle_{{\rm at}}$, the state of the system 
will evolve in the invariant subspace ${\cal H}_{L}$ with
$L=n_{0}+m_{0}+j$. For the field and/or atoms prepared 
initially in a superposition state, one should take into
account contributions from different subspaces.

The exact solution of the problem is obtained by the 
diagonalization of the interaction Hamiltonian (\ref{Ham}) in 
each of the invariant subspaces ${\cal H}_{L}$ involved 
\cite{TC68}. It is known \cite{TC68,WoBa70} that in the basis
$|n\rangle_{{\rm f}} |j,L-j-n\rangle_{{\rm at}}$ 
the Hamiltonian is given by a tridiagonal matrix with 
symmetric eigenvalues and the corresponding characteristic 
equation can be reduced to an algebraic equation of order 
$[{\rm dim}({\cal H}_{L})/2]$. 
Therefore, an analytical solution is possible when only 
invariant subspaces with ${\rm dim}({\cal H}_{L}) \leq 9$ are 
involved. However, already for ${\rm dim}({\cal H}_{L}) > 3$, 
analytical solutions are rather complicated 
\cite{Sen71,KoChSwMa92,KoChMa93}. 
Note that for ${\rm dim}({\cal H}_{L}) \geq 4$ the eigenvalues
are not equidistant, so the time evolution is not periodic even
within a single subspace.
Semiclassical approximate solutions were proposed 
\cite{BoPr70,KuMe80,KaHu81} which give the time evolution of 
systems governed by a trilinear Hamiltonian of type 
(\ref{Ham}) in terms of elliptical functions.
A perturbative analytical approach to the problem with weak
fields was developed by Kozierowski and co-workers 
\cite{KoChSwMa92,KoChMa93,KoMaCh90,KoChMa92}.
In the present work we use the exact solution based on the 
numerical diagonalization of the interaction Hamiltonian 
(\ref{Ham}).

\section{Collapses and revivals}

We study the temporal behavior of the atom-field system in 
the Dicke model for the cavity field prepared initially in 
the coherent state $|\alpha\rangle_{{\rm f}}$:
\begin{equation}
|\alpha\rangle_{{\rm f}} = e^{-|\alpha|^{2}/2} 
\sum_{n=0}^{\infty} \frac{\alpha^{n}}{\sqrt{n!}} 
|n\rangle_{{\rm f}} .
\end{equation}
Without loss of generality we consider only real values of
$\alpha$. The initial atomic state is supposed to be one
of the Dicke states $|j,m\rangle_{{\rm at}}$ (recall that
$j=N/2$). Two possibilities which are frequently considered 
in the literature are the fully-excited state 
$|j,j\rangle_{{\rm at}}$ and the ground state 
$|j,-j\rangle_{{\rm at}}$. We are also interested in the
half-excited Dicke state $|j,0\rangle_{{\rm at}}$.

\begin{figure}[htbp]
\epsfxsize=0.45\textwidth
\centerline{\epsffile{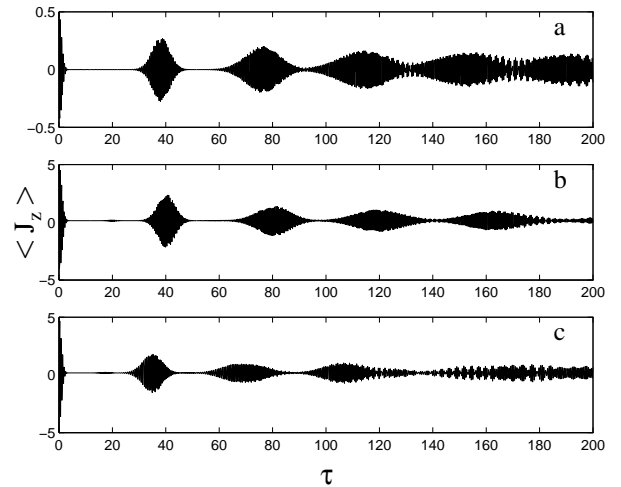}}
\vspace*{1mm}
\caption{The atomic inversion $\langle J_{z} \rangle$ 
versus the scaled time $\tau = g t$ for the initial coherent 
field state with $\bar{n} = 36$ interacting with (a) one excited
atom (the Jaynes-Cummings model) and with $10$ atoms prepared in 
(b) the fully-excited state $|j,j\rangle_{{\rm at}}$ and
(c) the ground state $|j,-j\rangle_{{\rm at}}$.}
\label{cr1}
\end{figure}

Since the coherent field state is a superposition of many
Fock states $|n\rangle_{{\rm f}}$, the invariant subspaces 
${\cal H}_{L}$ with different values of $L=n+m+j$ contribute 
to the evolution. The temporal behavior of the atomic 
inversion $\langle J_{z} \rangle$ is given by the
sum of the appropriately weighted atomic responses to each
Fock state. 
(The mean photon number $\langle a^{\dagger} a \rangle$ is 
connected with the atomic inversion by the fact that 
$\langle L \rangle = \langle a^{\dagger} a \rangle +
\langle J_{z} \rangle + N/2$ is a constant.)
In the resonant Jaynes-Cummings model ($N=1$) with initially 
unexcited ($m=-\frac{1}{2}$) or excited ($m=\frac{1}{2}$) atom, 
one obtains \cite{EbNaSM}
\begin{equation}
\langle J_{z} \rangle = m \sum_{n=0}^{\infty} P_{n} 
\cos \Omega_{n,m} \tau  ,
\end{equation}
where $\tau = gt$ is the scaled time, 
\begin{equation}
\Omega_{n,m} = 2\sqrt{n+m+1/2}
\end{equation}
is the scaled Rabi frequency corresponding to each subspace, 
and $P_{n}$ is the photon-number distribution. For the initial 
coherent state, $P_{n} = \exp(-\bar{n}) \bar{n}^n/n!$ is the 
Poissonian distribution and $\bar{n} = |\alpha|^2$ is the initial 
mean photon number. Then, due to the property of this distribution, 
the most regular dynamics occurs for large values of the 
initial mean photon number $\bar{n}$.
Contributions corresponding to different $n$'s interfere in such 
a manner that they initially go out of phase, after that acquire 
a common phase, and this process is repeated resulting in a 
series of collapses and revivals, as shown in Fig.\ \ref{cr1}(a).
The revival time $\tau_{R}$ can be estimated using the condition 
\cite{EbNaSM,Milb84}
\begin{equation}
\tau_{R}(\Omega_{\bar{n}+1,m}-\Omega_{\bar{n},m}) = 2\pi ,
\end{equation}
which gives
$\tau_{R} = 2\pi \sqrt{\bar{n}+m+1/2}$.

One should expect a similar behavior also for the Dicke model
in the strong-field domain $\bar{n} > N$ 
\cite{BaKn84,DroJex,ChKlSM94,ChKo96}. Of course,
the collapses and revivals related to the photon-number 
distribution would be modified by the collective atomic effects
due to the fact that the eigenvalues of the interaction 
Hamiltonian are not equidistant. In the strong-field limit
$\bar{n} \gg N$ the anharmonic corrections to the eigenvalues
become small, and one should expect a quite regular behavior
similar to that of the resonant Jaynes-Cummings model.
However, we will see that there exist reasons for an irregular 
behavior that are much more important than just the anharmonicity 
of the eigenvalues. In fact, the initial atomic state determines 
how important will be various factors leading to irregularities 
in the behavior of the system. 

\begin{figure}[htbp]
\epsfxsize=0.45\textwidth
\centerline{\epsffile{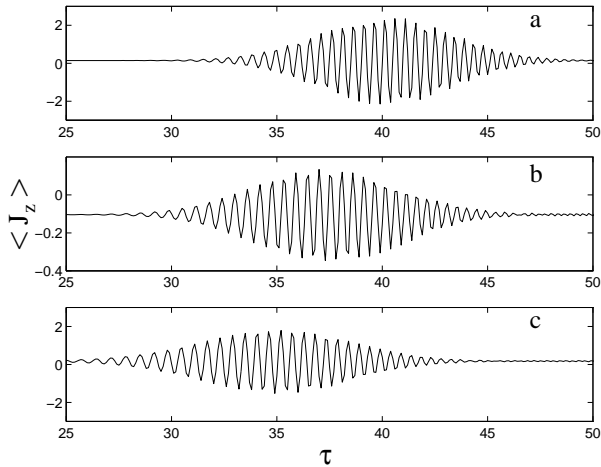}}
\vspace*{1mm}
\caption{The atomic inversion $\langle J_{z} \rangle$ 
versus the scaled time $\tau = g t$ for the initial coherent 
field state with $\bar{n} = 36$ interacting with with $10$ 
atoms prepared in 
(a) the fully-excited state $|j,j\rangle_{{\rm at}}$, 
(b) the half-excited Dicke state $|j,0\rangle_{{\rm at}}$, and
(c) the ground state $|j,-j\rangle_{{\rm at}}$.
The region of the first revival is shown, demonstrating the
dependence of the revival time on the initial atomic state.}
\label{cr2}
\end{figure}

We first consider the cases with the atoms prepared initially 
in the fully-excited and ground states. The temporal behavior 
of the atomic inversion $\langle J_{z} \rangle$ is shown in
Fig.\ \ref{cr1}(b,c) for the case of $N=10$ and $\bar{n}=36$, 
with atoms prepared in (b) the fully-excited state 
$|j,j\rangle_{{\rm at}}$ 
and (c) the ground state $|j,-j\rangle_{{\rm at}}$.
The first feature to note is the dependence of the 
revival time on the initial atomic state. Similarly to the 
Jaynes-Cummings model, we can estimate the revival time for 
the initial atomic state $|j,m\rangle_{{\rm at}}$ as
\begin{equation}
\tau_{R} = 2\pi \sqrt{\bar{n}+m+1/2} ,       \label{tauR}
\end{equation}
where we use the strong-field limit expression for the Rabi 
frequency (i.e., neglect the anharmonic corrections to the 
eigenvalues). Formula (\ref{tauR}) is in a good agreement 
with our numerical results, as demonstrated in Fig.\ \ref{cr2}.
The difference between the revival times for the fully-excited 
state and the ground state is particularly obvious when $N$ is 
not too small compared to $\bar{n}$. 
The second feature is that the spread of the revivals is 
determined only by the photon statistics of the initial field 
state \cite{Milb84} and does not depend on the value of $m$.
These two observations may explain why the temporal behavior 
loses its regularity for the initial ground state notably faster
than for the initial fully-excited state. Indeed, we see in
Fig.\ \ref{cr1}(b) four regular revivals for $m=j$, while in
Fig.\ \ref{cr1}(c) for $m=-j$ the behavior starts to be 
irregular already at the end of the third revival. 
The reason is that the spread of the revivals
increases with time and neighboring revivals start to overlap,
which leads to the loss of regularity. Since for $m=-j$ the 
revival time is shorter than for $m=j$, the overlapping of
the revivals and the corresponding irregularity occur earlier.

Another factor that leads to differences between the behavior 
of the system in the cases $m=j$ and $m=-j$ is the dependence 
of the anharmonic corrections to the eigenvalues on the value 
of $m$. In order to give a representative example, we considered 
the eigenvalues $h_{i}$ of the interaction Hamiltonian for the 
subspaces with $L=n+m+j$ for $n=36$, $N=10,20,30$ and $m=\pm j,0$. 
The ratio $f_{51} = h_{5}/h_{1}$ 
of the 5th and 1st eigenvalues was chosen as a parameter 
representing the anharmonicity (for equidistant eigenvalues
$f_{51}$ is exactly $5$). The numerical results are listed in 
Table \ref{T1}. We see that for given $N$ the value of $f_{51}$ 
increases with decrease of $m$, i.e., irregular effects related 
to the anharmonicity are most important for $m=-j$. For $m=j$ 
the value of $f_{51}$ decreases slightly with increase of $N$, 
but the overall contribution of the anharmonic corrections 
increases with $N$ just because there are more eigenvalues.
For $m=-j$ the value of $f_{51}$ increases with $N$, so the
importance of the anharmonic corrections here increases with 
$N$ much faster than for $m=j$. In particular, we see that 
the amplitude of the revivals for $m=-j$ is smaller than for  
$m=j$, and this effect becomes more pronounced as $N$ increases.
For $\bar{n}=36$, as $N$ increases from $2$ to $16$, the 
relative amplitude $A_{1}$ (the difference between the maximum 
and minimum values of $\langle J_{z} \rangle$ in the first 
revival, divided by $N$) decreases from $0.5459$ to $0.4039$ 
for $m=j$ and from $0.5362$ to $0.2112$ for $m=-j$. This effect 
occurs because the anharmonic corrections partially destroy the 
interference of the oscillating terms. Of course, as the initial 
mean photon number $\bar{n}$ increases, the behavior of the 
system becomes more regular. 

\begin{figure}[htbp]
\epsfxsize=0.45\textwidth
\centerline{\epsffile{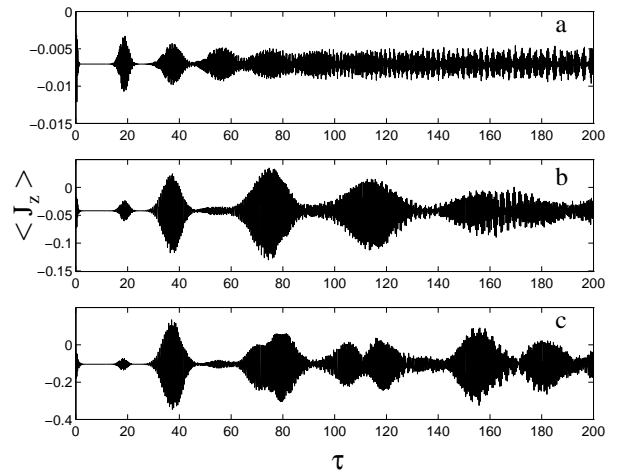}}
\vspace*{1mm}
\caption{The atomic inversion $\langle J_{z} \rangle$ 
versus the scaled time $\tau = g t$ for the initial coherent 
field state with $\bar{n} = 36$ interacting with $N$ atoms 
prepared in the half-excited state $|j,0\rangle_{{\rm at}}$:
(a) $N=2$, (b) $N=6$, (c) $N=10$.}
\label{cr3}
\end{figure}

Now we turn to the case when the atoms are prepared in the
half-excited Dicke state $|j,0\rangle_{{\rm at}}$ (for even
values of $N$).
The temporal behavior of the atomic inversion 
$\langle J_{z} \rangle$ is shown in Fig.\ \ref{cr3} for
$\bar{n} = 36$ and $N=2,6,10$.
Here we see a number of interesting phenomena. First of all, the 
amplitude of the Rabi oscillations is significantly suppressed. 
For $\bar{n}=36$, as $N$ increases from $2$ to $16$, the 
relative amplitude $A_{0}$ (the difference between the maximum 
and minimum values of $\langle J_{z} \rangle$ in the very 
beginning of the evolution, divided by $N$) increases linearly 
from $0.006$ to $0.027$, according to the empirical formula
\begin{equation}
A_{0} = 0.0015 (N+2) .
\end{equation}
For comparison, $A_{0}$ is about $0.9$ for $m=j$ 
and about $0.95$ for $m=-j$, being almost independent of $N$.

Another important feature is the appearance of half-time 
revivals, i.e., for $m=0$ the revival time is actually
$\tau_{R}/2$. In fact, tiny half-time revivals appear also 
for $m=\pm j$, but their amplitudes are much smaller than 
the amplitudes of the first revivals at $\tau_{R}$. 
As $N$ increases from $2$ to $16$, the relative amplitude 
$A_{1/2}$ (the difference between the maximum and minimum 
values of $\langle J_{z} \rangle$ in the half-time revival, 
divided by $N$) increases monotonically from $0.0018$ to
$0.0171$ for $m=j$ and from $0.0019$ to $0.0354$ for $m=-j$.
The ratio $A_{1/2}/A_{1}$ increases in the same range of
$N$ from $0.0033$ to $0.0424$ for $m=j$ and from $0.0035$
to $0.1677$ for $m=-j$. 
On the other hand, for $m=0$ the half-time revival and the 
first revival are, for small values of $N$, of the same order 
of magnitude. For $N=2$ we even find $A_{1/2}/A_{1}>1$.
In contradistinction to the cases $m=\pm j$, for $m=0$ the ratio 
$A_{1/2}/A_{1}$ decreases from $1.3427$ to $0.0473$ as $N$ 
increases from $2$ to $16$. In this range of $N$ the relative 
amplitude $A_{1/2}$ of the half-time revival increases slightly 
from 0.0037 to 0.0052, while the relative amplitude
$A_{1}$ of the first revival increases much faster: from 
$0.0028$ to $0.1093$. Starting from $N=4$ the amplitude of
the first revival exceeds that of the initial oscillations.
The ratio $A_{1}/A_{0}$ increases almost linearly from $0.4555$ 
to $4.0476$ as $N$ increases from $2$ to $16$. The value 
$\langle J_{z} \rangle_{C}$ of the atomic inversion during the 
collapse is always positive for $m=\pm j$. As $N$ increases 
from $2$ to $16$, $\langle J_{z} \rangle_{C}$ increases from
$0.0036$ to $0.3569$ for $m=j$ and from $0.0035$ to $0.5195$
for $m=-j$. On the other hand, for $m=0$ the value of
$\langle J_{z} \rangle_{C}$ is always negative and decreases
from $-0.0070$ to $-0.2433$ in the same range of $N$.
We found that for $m=0$ the value of $\langle J_{z} \rangle_{C}$
can be well approximated by the empirical formula
\begin{eqnarray}
& & \langle J_{z} \rangle_{C} = -\left( \frac{N
}{k_{1}\alpha-k_{2}} \right)^{z(\alpha)} , \nonumber \\
& & z(\alpha) = k_{3}-\exp[-(k_{4}\alpha+k_{5})] ,
\end{eqnarray}
where $k_{1}=7.45$, $k_{2}=11.16$, $k_{3}=1.773$, 
$k_{4}=0.328$, $k_{5}=1.681$. 
We also see that in general the dynamics in the case $m=0$ is 
much more irregular than for $m=\pm j$. All these observations 
show that the structure of the phenomenon of collapses and 
revivals is essentially different for the half-excited Dicke 
state relative to the fully-excited state or the ground state. 
Inspecting the eigenvalues (see Table \ref{T1}), we see that
the influence of the anharmonic corrections cannot explain
this principal difference.
As expected, the behavior becomes more regular with increase of
the mean photon number $\bar{n}$. For odd numbers of atoms
the half-excited Dicke state $|j,0\rangle_{{\rm at}}$ does not 
exist. However, for sufficiently large odd values of $N$ the 
Dicke state $|j,\frac{1}{2}\rangle_{{\rm at}}$ exhibits 
properties very similar to those of the half-excited state.

In order to explain the peculiar phenomena discussed above, we 
consider the exactly solvable case $N=2$. For the fully-excited 
state ($m=1$) we find
\begin{eqnarray}
\langle J_{z} \rangle & = & \sum_{n=0}^{\infty} P_{n} 
\frac{2}{\Omega_{n,1}^{4}} \left[ (n+3) - 
(n+1) \cos(2\Omega_{n,1}\tau) \right. \nonumber \\
& & \left. + 8(n+1)(n+2) \cos(\Omega_{n,1}\tau) 
\right] .  
\end{eqnarray}
Analogously, we obtain for the ground state ($m=-1$):
\begin{eqnarray}
\langle J_{z} \rangle & = & \sum_{n=0}^{\infty} P_{n} 
\frac{2}{\Omega_{n,-1}^{4}} \left[ (n-2) -  
n \cos(2\Omega_{n,-1}\tau) \right. \nonumber \\
& & \left. - 8n(n-1) \cos(\Omega_{n,-1}\tau) 
\right] , 
\end{eqnarray}
and for the half-excited Dicke state ($m=0$):
\begin{equation}
\langle J_{z} \rangle = - \sum_{n=0}^{\infty} P_{n}
\frac{1}{\Omega_{n,0}^{2}} \left[1-\cos(2\Omega_{n,0}\tau)
\right] . 
\label{popinv0}
\end{equation}
Now it is clear why the oscillations are strongly suppressed
for $m=0$. Here both the constant and oscillatory terms have the
prefactor $P_{n}/(4n+2)$, while for $m=\pm j$ in addition to
the constant and oscillatory terms with prefactors of the order
$P_{n}/n$ there exists an oscillatory term with a prefactor of
the order $P_{n}$. For $\bar{n}=36$ the suppression is by 
two orders of magnitude. For $m=0$ the oscillatory term has the
frequency $2\Omega_{n,0}=2(n+1/2)^{1/2}$, and the effective
revival time is $\tau_{R}/2 = \pi (\bar{n}+1/2)^{1/2}$. 
On the other hand, for $m=\pm j$ the leading oscillatory term 
with the prefactor of the order $P_{n}$ has the frequency 
$\Omega_{n,\pm 1}$, associated with the revival time $\tau_{R}$, 
while the smaller oscillatory term with the prefactor of the 
order $P_{n}/n$ has the double frequency, leading to the
half-time revivals. This explains why for $m=\pm j$ the 
half-time revival is much smaller than the first revival
while for $m=0$ both types of revivals are of the same 
order of magnitude. The more irregular dynamics in the case
$m=0$ can be explained by two reasons. First, the nonlinearity
of the Rabi frequency ($\sim \sqrt{n}$) is less important for 
larger values of $n$, so the dynamics is more regular when the
main contribution comes from larger $n$'s. For $m=\pm j$ the
prefactors of the leading terms are Poissonian, so for large
enough $\bar{n}$, the main contribution will come from the
high-frequency terms, resulting in a regular behavior. 
However, for $m=0$ the prefactors are $P_{n}/(4n+2)$, so lower
frequencies also contribute, which results in a less regular
behavior. The second reason is that the revivals start to 
overlap much earlier if the revival time is halved. The same
reason also leads to an additional irregularity in the case
$m=-j$ when the number of atoms is relatively large and the 
half-time revivals are not too small.

The suppression of the revival amplitude for the half-excited
Dicke state is similar to the trapping phenomenon which
occurs for a single atom prepared in the equally weighted
superposition state $2^{-1/2}(|+\rangle \pm |-\rangle)$ 
(see Ref.~\cite{ZaZu89}). 
In the latter case the population inversion is given by
\begin{eqnarray}
\langle J_{z} \rangle & = & \frac{1}{2} \sum_{n=0}^{\infty} P_{n}
[ \cos^{2}(\sqrt{n+1}\tau) \nonumber \\
& &  + \frac{\bar{n}}{n+1} \sin^{2}(\sqrt{n+1}\tau) ] 
-  \frac{1}{2} . 
\end{eqnarray}
For large $\bar{n}$, the Poissonian distribution is sharply
peaked around $\bar{n}$, and the two terms in the sum almost
add up to $1$. The remaining oscillating term has the
prefactor of the order $P_{n}/n$, so the amplitude of the
Rabi oscillations is reduced by the factor of the order
$1/\bar{n}$. As explained above, in such a situation
the dynamics is less regular than for the case of
initially unexcited or fully excited atom. However, the
revival time is not halfed for the single atom in the 
equally weighted superposition state. In the single-atom case 
the population trapping occurs due to the destructive 
interference between the contributions of the two levels, 
while for the two-atom half-excited Dicke state this
phenomenon can be explained by the destructive interference 
between the contributions of the two correlated atoms.

\section{Squeezing of the radiation field}

The coherent radiation field interacting with atoms can acquire 
interesting nonclassical properties such as sub-Poissonian
photon statistics and squeezing. In the present paper we
focus on the important quantum phenomenon of squeezing.
The initially coherent cavity field can be squeezed when it 
interacts with a single atom \cite{MeZu82,ShFLKAl87}, and
squeezing in the revival-time regime can be very strong for
large intensities of the field \cite{KuMa88,WoGB93}.
Butler and Drummond discovered \cite{BuDr86} that in the Dicke 
model collective atomic effects can improve squeezing obtained
for short interaction times, compared to the single-atom case. 
Squeezing in the Dicke model was also considered recently in a 
number of works \cite{Seke:sq,RSKC97,KoCh97}. Here we present 
a detailed study of squeezing of the initially coherent cavity 
field, comparing between the ground, fully-excited and 
half-excited initial atomic states.

The field quadratures, $q=(a^{\dagger}+a)/\sqrt{2}$ and 
$p= i (a^{\dagger}-a)/\sqrt{2}$, satisfy the canonical 
commutation relation $[q,p] = i$. Then their uncertainties 
satisfy the Heisenberg relation: 
$\Delta q \Delta p \geq 1/2$, 
where $(\Delta q)^{2} = \langle q^{2} \rangle - 
\langle q \rangle^{2}$ and similarly for $(\Delta p)^{2}$. 
For the coherent field state (in particular, for the vacuum
state) the uncertainties are equal, 
$\Delta q_{0} = \Delta p_{0} = 1/\sqrt{2}$, and an equality 
is achieved in the Heisenberg uncertainty relation.
A field state is called squeezed, if the uncertainty of one
of the quadratures is below the vacuum level, i.e.,
$\Delta q < 1/\sqrt{2}$ or $\Delta p < 1/\sqrt{2}$.
Here we consider the squeezing parameter
\begin{equation}
\xi = \frac{\Delta q}{\Delta q_{0}} = \sqrt{2} \Delta q ,
\end{equation}
and search for $\xi < 1$ that manifests field squeezing.

\subsection{Squeezing for the fully-excited and ground states}

\subsubsection{The long-time behavior}

We first consider the temporal behavior of the squeezing 
parameter $\xi$ on the long-time scale, for the atoms prepared
initially in the fully-excited and ground states ($m = \pm j$).

\begin{figure}[htbp]
\epsfxsize=0.45\textwidth
\centerline{\epsffile{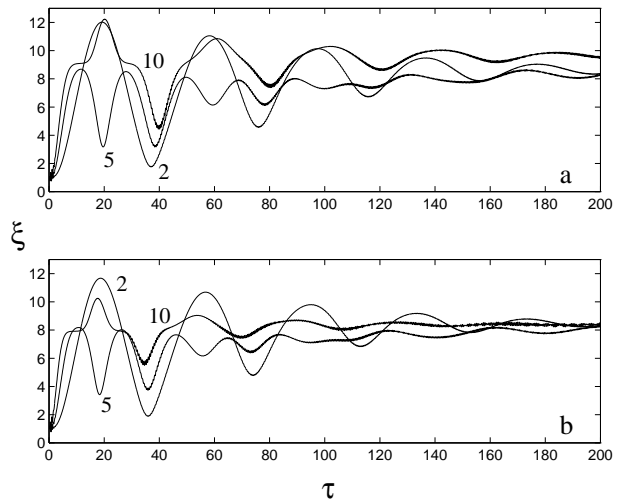}}
\vspace*{1mm}
\caption{The squeezing parameter $\xi$ versus the scaled time 
$\tau = g t$ for the initial coherent field state with 
$\bar{n} = 36$ interacting with $N$ atoms ($N=2,5,10$) 
prepared in 
(a) the fully-excited state $|j,j\rangle_{{\rm at}}$ and
(b) the ground state $|j,-j\rangle_{{\rm at}}$.}
\label{sqlt}
\end{figure}

As demonstrated in Fig.\ \ref{sqlt}, this long-time behavior
is fully correlated with the collapses and revivals of the Rabi 
oscillations. For even values of $N$, the squeezing parameter 
$\xi$ oscillates achieving its minima at integer multiples of 
$\tau_{R}$ and maxima at half-odd multiples of $\tau_{R}$. 
On the other hand, for odd values of $N$, we see that $\xi$ 
has minima at both half-odd and integer multiples of $\tau_{R}$.
This difference between even and odd values of $N$ can be 
explained by the fact that squeezing depends on two-photon
transitions, when a pair of photons is simultaneously absorbed 
or emitted by a pair of atoms. The oscillations of $\xi$ decay
with time and abandon their regular form. This decay is  
correlated with the loss of regularity in the behavior of the 
atomic inversion, caused by the overlaps of neighboring 
revivals. The behavior of $\xi$ is rather similar for $m=j$
and $m=-j$, but in the latter case the decay of the 
oscillations and the irregularity occur earlier, as the revival
time is shorter and the overlaps of revivals begin earlier.
Also, for $m=-j$ the decay rate of the squeezing oscillations 
increases rapidly with $N$, while for $m=j$ the dependence of 
the decay rate on $N$ is less pronounced.
Inspecting the minima of $\xi$ in Fig.\ \ref{sqlt}, we clearly
see that their period becomes longer for $m=j$ and shorter for
$m=-j$ as $N$ increases, in full accordance with the formula
(\ref{tauR}) for the revival time. 

For $m = \pm j$, the minima of $\xi$ on the long-time scale 
are above the vacuum level, i.e., $\xi > 1$. However, squeezing 
is achieved at short times, soon after the beginning of the 
interaction. In this region the squeezing parameter exhibits 
fast oscillations, with $\xi$ falling below $1$. Therefore, for
$m = \pm j$ we will focus on the behavior of squeezing 
on the short-time scale.

\subsubsection{Squeezing on the short-time scale: 
The fully-excited state}

We first consider the case of the initial fully-excited atomic
state ($m=j$). Squeezing is achieved for $\tau < 2$ and appears
at shorter times for larger values of $N$. 

\begin{figure}[htbp]
\epsfxsize=0.45\textwidth
\centerline{\epsffile{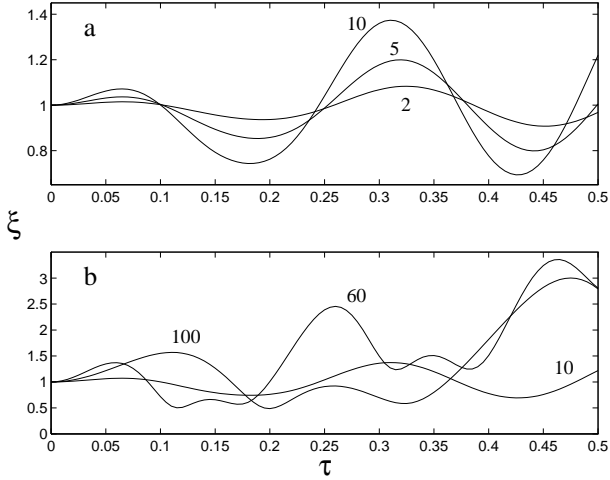}}
\vspace*{1mm}
\caption{The squeezing parameter $\xi$ versus the scaled time 
$\tau = g t$ for the initial coherent field state with 
$\bar{n} = 36$ interacting with $N$ atoms prepared in 
the fully-excited state:
(a) $N=2,5,10$,
(b) $N=10,60,100$.}
\label{sqpst1}
\end{figure}

\begin{figure}[htbp]
\epsfxsize=0.45\textwidth
\centerline{\epsffile{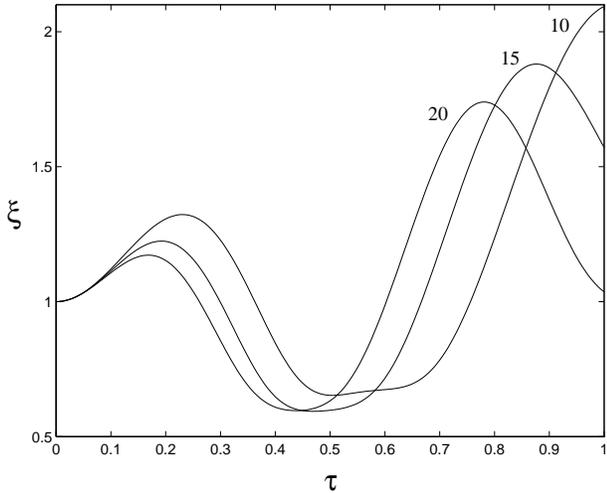}}
\vspace*{1mm}
\caption{The squeezing parameter $\xi$ versus the scaled time 
$\tau = g t$ for the initial coherent field state with 
$\bar{n} = 10,15,20$ interacting with $14$ atoms prepared in 
the fully-excited state.}
\label{sqpst2}
\end{figure}

In Fig.\ \ref{sqpst1} we see the short-time behavior of the 
squeezing parameter $\xi$ for $\bar{n} = 36$ and various values 
of $N$. For relatively small values of $N$ ($N \leq 10$), $\xi$
exhibits quite regular oscillations whose amplitude increases
with $N$. However, for $N \sim \bar{n}$ and larger, the
oscillations of $\xi$ become irregular. In Fig.\ \ref{sqpst2}
the short-time behavior of $\xi$ is shown for $N=14$ and
various values of $\bar{n}$. The oscillations of $\xi$ become
more regular as $\bar{n}$ increases. 

It is interesting to investigate how the minimum value $\xi_{m}$ 
(i.e., the maximum of squeezing), achieved during the time 
evolution, depends on $N$ and $\bar{n}$. 

\begin{figure}[htbp]
\epsfxsize=0.45\textwidth
\centerline{\epsffile{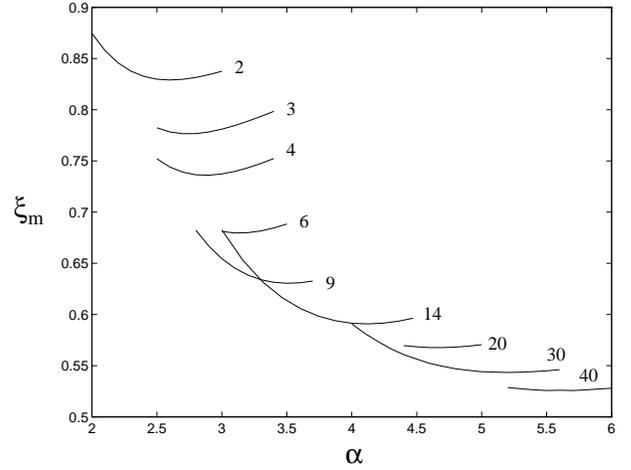}}
\vspace*{1mm}
\caption{The minimum value $\xi_{m}$ of the squeezing parameter 
versus the coherent amplitude $\alpha$ for various values of $N$.}
\label{sqpst3}
\end{figure}
\vspace*{3mm}
\begin{figure}[htbp]
\epsfxsize=0.45\textwidth
\centerline{\epsffile{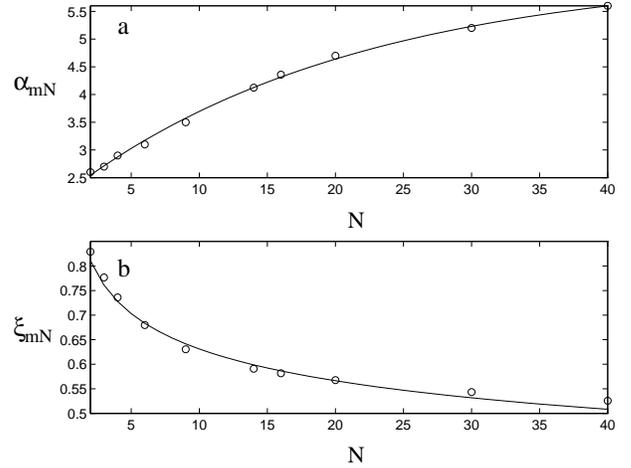}}
\vspace*{1mm}
\caption{(a) The value $\alpha_{mN}$ of the coherent amplitude,
for which the squeezing parameter is minimized, versus $N$:
numerical results (circles) and empirical fitting of Eq.\
(\protect\ref{alphamN}) (line);
(b) the corresponding value $\xi_{mN}$ of the squeezing 
parameter versus $N$: numerical results (circles) and 
empirical fitting of Eq.\ (\protect\ref{xmN}) (line).}
\label{sqpst3d}
\end{figure}

We first consider the 
dependence of $\xi_{m}$ on $\alpha = \sqrt{\bar{n}}$ for given 
$N$. As shown in Fig.\ \ref{sqpst3}, $\xi_{m}$ has a minimum as 
a function of $\alpha$, i.e., for given $N$ there exists a value 
$\alpha_{mN}$ for which the minimum value $\xi_{mN}$ of the
squeezing parameter is achieved. Figure \ref{sqpst3d} shows
that $\alpha_{mN}$ increases and $\xi_{mN}$ decreases (squeezing 
improves) as $N$ increases. We found that the dependence of 
$\alpha_{mN}$ and $\xi_{mN}$ on $N$ at the considered range 
can be well approximated by the following empirical formulas:
\begin{eqnarray}
& & \alpha_{mN} = a - b e^{-c N} , \label{alphamN} \\
& & \xi_{mN} = r N^{-s} , \label{xmN}
\end{eqnarray}
where $a=6.21$, $b=4.03$, $c=0.0471$, $r=0.909$, $s=0.156$.

\begin{figure}[htbp]
\epsfxsize=0.45\textwidth
\centerline{\epsffile{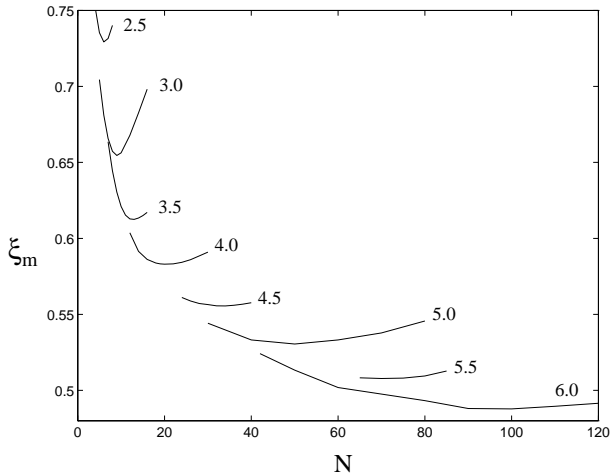}}
\vspace*{1mm}
\caption{The minimum value $\xi_{m}$ of the squeezing parameter 
versus $N$ for various values of $\alpha$.}
\label{sqpst4}
\end{figure}
\vspace*{-3mm}
\begin{figure}[htbp]
\epsfxsize=0.45\textwidth
\centerline{\epsffile{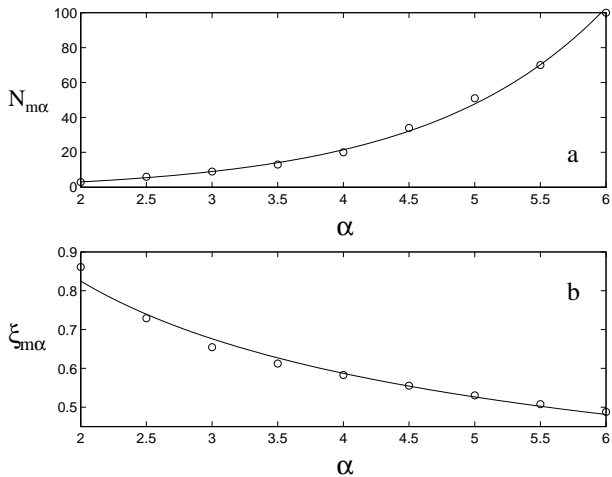}}
\vspace*{1mm}
\caption{(a) The value $N_{m\alpha}$ of the number of atoms,
for which the squeezing parameter is minimized, versus $\alpha$:
numerical results (circles) and empirical fitting of Eq.\
(\protect\ref{Nmalpha}) (line);
(b) the corresponding value $\xi_{m\alpha}$ of the squeezing 
parameter versus $\alpha$: numerical results (circles) and 
empirical fitting of Eq.\ (\protect\ref{xmalpha}) (line).}
\label{sqpst4d}
\end{figure}

We next consider the dependence of $\xi_{m}$ on $N$ for given 
$\alpha$. As shown in Fig.\ \ref{sqpst4}, $\xi_{m}$ has a 
minimum as a function of $N$, i.e., for given $\alpha$ there 
exists a value $N_{m\alpha}$ for which the minimum value 
$\xi_{m\alpha}$ of the squeezing parameter is achieved. 
Figure \ref{sqpst4d} shows that $N_{m\alpha}$ increases and 
$\xi_{m\alpha}$ decreases (squeezing improves) as $\alpha$ 
increases. We found again that the dependence of $N_{m\alpha}$ 
and $\xi_{m\alpha}$ on $\alpha$ at the considered range can be 
well approximated by the following empirical formulas:
\begin{eqnarray}
& & N_{m\alpha} = -a + b e^{c \alpha} , \label{Nmalpha} \\
& & \xi_{m\alpha} = r \alpha^{-s} , \label{xmalpha}
\end{eqnarray}
where $a=2.30$, $b=1.216$, $c=0.743$, $r=1.159$, $s=0.490$.
Note that the relations (\ref{alphamN}) and (\ref{Nmalpha})
are not the inverse of each other. This fact can be easily 
understood, if one imagine $\xi_{m}$ as a two-dimensional
function of $N$ and $\alpha$. When taking a section of 
$\xi_{m}$ along the $\alpha$ axis (i.e., for a given $N$),
one will find the minimum for a certain value of $\alpha$.
However, when fixing this value of $\alpha$ and going along 
the $N$ axis, a minimum will be found, in general, for a 
different $N$.

Finally, we would like to compare our numerical results with
approximate analytical expressions derived by Retamal 
{\em et al.} \cite{RSKC97} in the strong-field limit 
$\bar{n} \gg N$. They found \cite{RSKC97} the following 
expression for the minimum of the squeezing parameter,
achieved during the time evolution, 
\begin{equation}
\xi_{m} = \left( 1 - a \frac{N}{\alpha} + 
b \frac{N^2}{8\alpha^2} \right)^{1/2} ,
\label{Retappr}
\end{equation}
where $a=1/\sqrt{e} \approx 0.606$ and 
$b=1+a-a^{2}-a^{4} \approx 1.103$. Then, for given $\alpha$,
the squeezing parameter will be minimized by 
\begin{equation}
N_{m\alpha} = \frac{4a}{b} \alpha \approx 2.2 \alpha . 
\label{Nan}
\end{equation}

\vspace*{-4mm}
\begin{figure}[htbp]
\epsfxsize=0.45\textwidth
\centerline{\epsffile{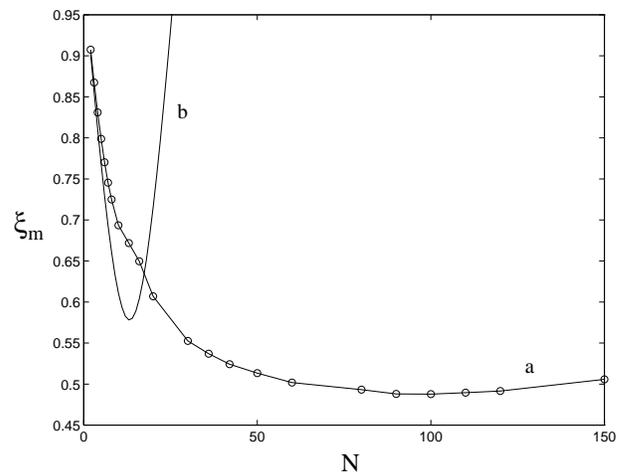}}
\vspace*{1mm}
\caption{The minimum value $\xi_{m}$ of the squeezing parameter 
versus $N$ for $\alpha = 6$: (a) numerical result,
(b) approximate analytical formula (\protect\ref{Retappr}).}
\label{sqpst5}
\end{figure}

In Fig.\ \ref{sqpst5} we compare our numerical results for 
$\xi_{m}$ with the approximate formula (\ref{Retappr}) for 
$\alpha = 6$ ($\bar{n} = 36$). A good agreement is found only 
for very small values of $N$, while the values of $N_{m\alpha}$ 
are absolutely different ($N_{m\alpha} \approx 100$ for our 
numerical calculations and $N_{m\alpha} \approx 13$ for the 
analytical approximation).
This discrepancy can be explained by the fact that the numerical
values of $N_{m\alpha}$ shown in Fig.\ \ref{sqpst4d} do not
satisfy the strong-field condition $\bar{n} \gg N$. For the
considered range of $\bar{n}$ ($\leq 36$), the approximate
solution (\ref{Retappr}) is not valid. From Eq.\ (\ref{Nan})
one can see that the strong-field condition will be satisfied
for the optimal value $N_{m\alpha}$ when 
$N_{m\alpha}/\bar{n} \approx 2.2/\alpha \ll 1$. This means that 
Eq.\ (\ref{Nan}) gives a true value of $N_{m\alpha}$ only for 
$\bar{n} \sim 10^{4}$ or more. According to Eq.\ (\ref{Retappr}), 
the absolute minimum of the squeezing parameter that can be 
achieved in the strong-field limit is $\xi \approx 0.58$. 
For $\bar{n}=36$ (the maximum $\bar{n}$ we considered), the best 
value of squeezing is $\xi \approx 0.49$ and it is achieved for
$N_{m\alpha} \approx 100$.

\subsubsection{Squeezing on the short-time scale: 
The ground state}

We next consider the case when the initial atomic state is the 
ground state ($m=-j$). In this case we can distinguish three 
different regimes: the strong-field regime ($\bar{n} \gg N$), 
the weak-field regime ($\bar{n} \ll N$), and the intermediate 
regime ($\bar{n} \sim N$). 

\begin{figure}[htbp]
\epsfxsize=0.45\textwidth
\centerline{\epsffile{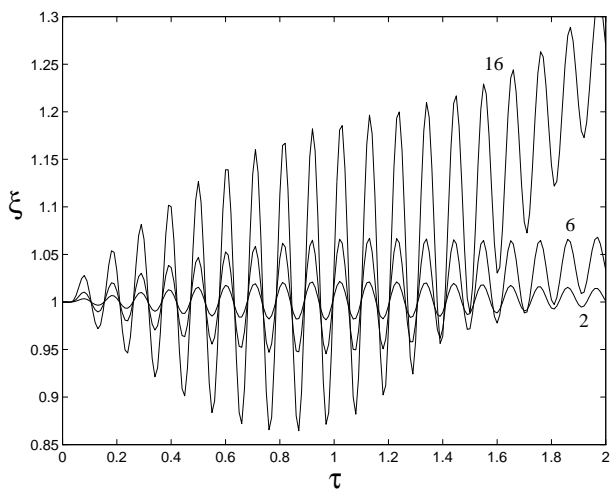}}
\vspace*{1mm}
\caption{The squeezing parameter $\xi$ versus the scaled time 
$\tau = g t$ for the initial coherent field state with 
$\alpha = 30$ interacting with $N$ atoms ($N = 2,6,16$)
prepared in the ground state.}
\label{sqmst1}
\end{figure}

\begin{figure}[htbp]
\epsfxsize=0.45\textwidth
\centerline{\epsffile{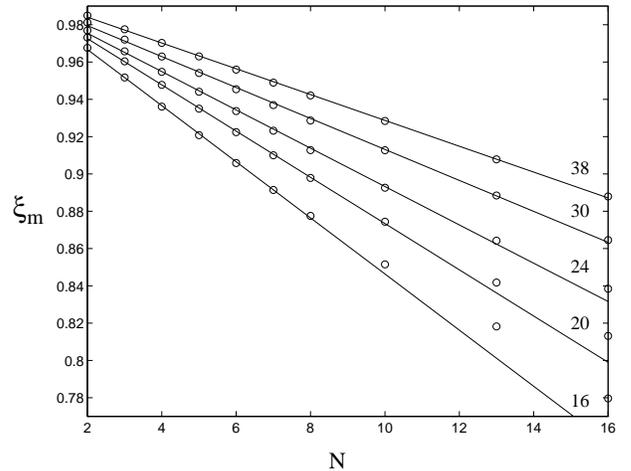}}
\vspace*{1mm}
\caption{The minimum value $\xi_{m}$ of the squeezing parameter 
versus $N$ for various values of $\alpha$: numerical results
(circles) and linear fitting of Eq.\ (\protect\ref{ximground}) (line).
The fitting is good in the strong-field regime $\bar{n} \gg N$.}
\label{sqmst2}
\end{figure}

The short-time behavior of the squeezing parameter $\xi$ in 
the strong-field regime is shown in Fig.\ \ref{sqmst1}
(for $\alpha = 30$ and $N=2,6,16$). 
We see that $\xi$ exhibits fast regular oscillations, whose 
frequency is almost independent of $N$ (this is just the 
strong-field Rabi frequency 
$\Omega_{\bar{n},-j}^{(s)} = 2\sqrt{\bar{n}-N/2+1/2}$, 
and the dependence on $N$ is very weak because $\bar{n}$ is 
very large), but the amplitude increases with $N$. Therefore, 
the minimum value $\xi_{m}$ of the squeezing parameter decreases 
as $N$ increases. A similar behavior is found also for the case 
$m=j$ in the strong-field regime. An interesting feature of 
this regime is that the minimum of $\xi$ is achieved after a 
relatively large number of oscillations (the same time for 
different values of $N$), while in the intermediate regime the 
minimum of $\xi$ is achieved, as a rule, in the first or second 
oscillation. As can be seen from Fig.\ \ref{sqmst2}, in the 
strong-field regime the decrease of $\xi_{m}$ is linear and can 
be well approximated by the empirical formula
\begin{equation}
\xi_{m} = - f(\alpha) N + k ,
\label{ximground}
\end{equation}
where the slope $f(\alpha)$ is a monotonically decreasing
function of $\alpha$ and the free term $k \approx 1$ is 
independent of $\alpha$.

\vspace*{-5mm}
\vspace*{-5mm}
\begin{figure}[htbp]
\epsfxsize=0.45\textwidth
\centerline{\epsffile{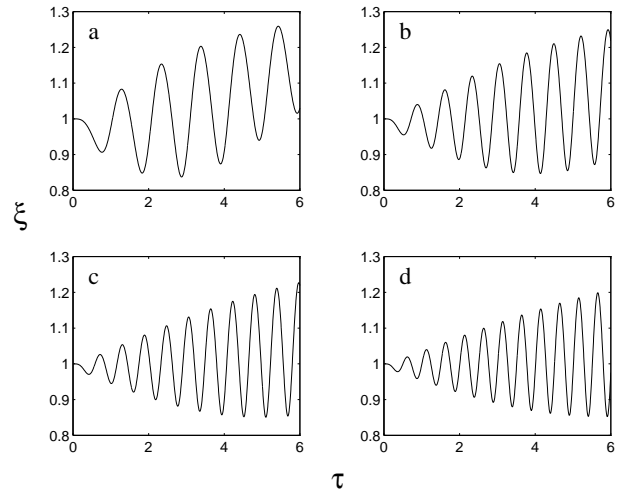}}
\vspace*{1mm}
\caption{The squeezing parameter $\xi$ versus the scaled time 
$\tau = g t$ for the initial coherent field state with 
$\alpha = 1$ interacting with $N$ atoms prepared in the ground 
state:
(a) $N=10$, (b) $N=20$, (c) $N=30$, (d) $N=40$.}
\label{sqmst3}
\end{figure}

It is well known \cite{KoChSwMa92,KoChMa93,KoMaCh90,KoChMa92}
that in the weak-field regime the behavior of the system is
very regular (note that for the case of the fully-excited 
initial state the weak-field regime actually does not exist). 
In Fig.\ \ref{sqmst3} we see the evolution of the squeezing 
parameter $\xi$ for relatively short times
(for $\alpha =1$ and $N=10,20,30,40$). The oscillations of 
$\xi$ are quite regular and their frequency increases with
$N$ (this is just the weak-field Rabi frequency 
$\Omega_{\bar{n},-j}^{(w)} = 2\sqrt{N-\bar{n}/2+1/2}$),
while the amplitude does not change significantly.
The value of squeezing achieved in this limit is rather
modest (as a rule, $\xi$ does not decrease below $0.85$).
A perturbative analytical approach to the Dicke model 
dynamics in the weak-field regime was developed by 
Kozierowski and co-workers 
\cite{KoChSwMa92,KoChMa93,KoMaCh90,KoChMa92} and used
for the study of squeezing in Ref.\ \cite{KoCh97}.

\begin{figure}[htbp]
\epsfxsize=0.45\textwidth
\centerline{\epsffile{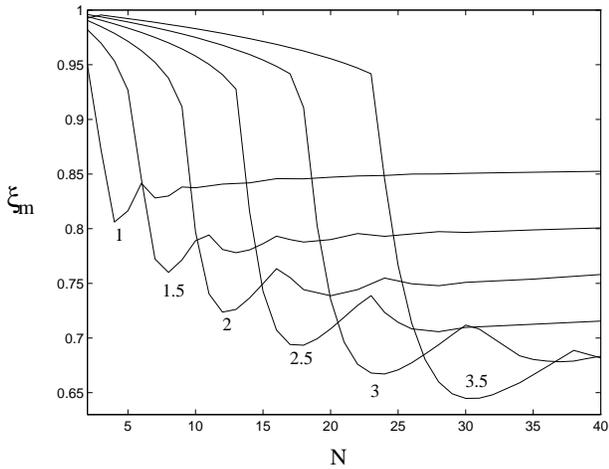}}
\vspace*{1mm}
\caption{The minimum value $\xi_{m}$ of the squeezing parameter 
versus $N$ for various values of $\alpha$.}
\label{sqmst4}
\end{figure}

We focus our attention on properties of squeezing in the
intermediate regime, where no analytical approximation can be
used. We study the minimum value $\xi_{m}$, achieved by the
squeezing parameter during the time evolution, for $N$ in the
range between $2$ and $40$ and for $\alpha$ between $1.0$ and
$6.5$. The results are presented in Fig. \ref{sqmst4}, 
where $\xi_{m}$ is plotted versus $N$ for various values of 
$\alpha$. The typical behavior for given $\alpha$ is as follows. 
Initially, $\xi_{m}$ decreases slowly with $N$, but then steeply 
sinks down and acquires a minimum at a certain value of $N$.
After the minimum, $\xi_{m}$ slightly oscillates and then
saturates for large $N$'s at an almost constant value. 
The region of linear decrease, occurring in the strong-field 
regime, appears only for $\alpha \geq 4$. This leads to an
additional minimum at small values of $N$, but it is less
pronounced than the main minimum at larger $N$'s.

\begin{figure}[htbp]
\epsfxsize=0.45\textwidth
\centerline{\epsffile{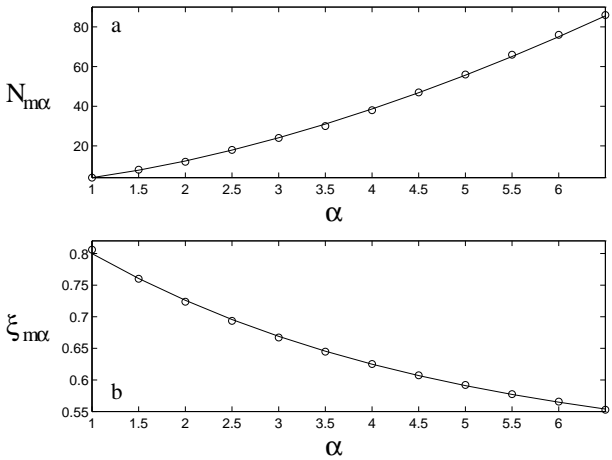}}
\vspace*{1mm}
\caption{(a) The value $N_{m\alpha}$ of the number of atoms,
for which the squeezing parameter is minimized, versus $\alpha$:
numerical results (circles) and empirical fitting of Eq.\
(\protect\ref{Nmamj}) (line);
(b) the corresponding value $\xi_{m\alpha}$ of the squeezing 
parameter versus $\alpha$: numerical results (circles) and 
empirical fitting of Eq.\ (\protect\ref{xmamj}) (line).}
\label{sqmst5}
\end{figure}

The value $N_{m\alpha}$, which gives optimal squeezing for given 
$\alpha$, and the corresponding minimum value $\xi_{m\alpha}$ of 
the squeezing parameter are shown in Fig.\ \ref{sqmst5} as 
functions of $\alpha$. As $\alpha$ increases, squeezing improves 
($\xi_{m\alpha}$ decreases) and the minimum occurs at larger
$N_{m\alpha}$. The dependence of $N_{m\alpha}$ and $\xi_{m\alpha}$ 
on $\alpha$ at the considered range can be well approximated by 
the following empirical formulas:
\begin{eqnarray}
& & N_{m\alpha} = k \alpha^{\gamma} , \label{Nmamj} \\
& & \xi_{m\alpha} = x + y e^{-z \alpha} , \label{xmamj}
\end{eqnarray}
where $k=1.382$, $\gamma=1.639$, $x=0.476$, $y=0.420$, $z=0.259$.
The largest value of $\alpha$ we considered is $6.5$. If we 
assume that Eq.\ (\ref{xmamj}) remains valid for arbitrarily 
large $\alpha$, then the absolute minimum of squeezing achievable 
with unexcited atoms is $\xi \approx 0.476$.

\subsection{Squeezing for the half-excited state 
and the factorization approximation}

\begin{figure}[htbp]
\epsfxsize=0.45\textwidth
\centerline{\epsffile{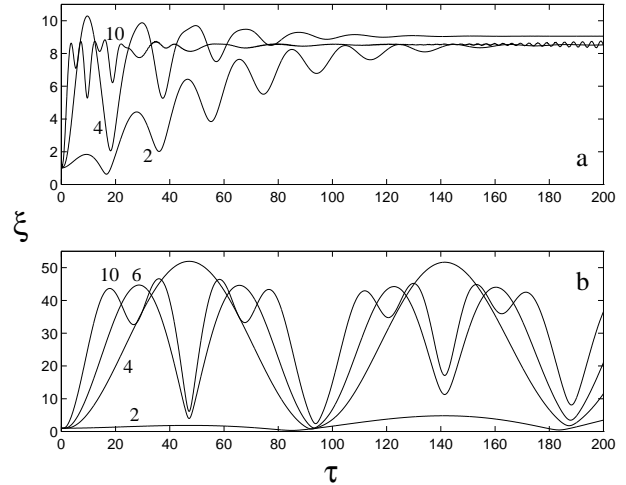}}
\vspace*{1mm}
\caption{The squeezing parameter $\xi$ versus the scaled time 
$\tau = g t$ for the initial coherent field state interacting 
with $N$ atoms prepared in the half-excited Dicke state:
(a) $\bar{n} = 36$, $N=2,4,10$,
(b) $\bar{n} = 900$, $N=2,4,6,10$.}
\label{sq01}
\end{figure}

Finally, we discuss the case when the initial atomic state is 
the half-excited Dicke state ($m=0$). The temporal behavior of 
the squeezing parameter $\xi$ on the long-time scale is shown 
in Fig.\ \ref{sq01}. We see that there is a strong correlation 
between the collapses and revivals of the atomic inversion and 
the behavior of squeezing. 

For $\bar{n} = 36$ and $N=2,4$, we see in Fig.\ \ref{sq01}(a) 
that $\xi$ oscillates with a period of order $\tau_{R}/2$ 
(recall that for $m=0$ we find relatively strong half-time 
revivals). However, after a number of periods, the amplitude 
of the oscillations decays and $\xi$ becomes nearly constant. 
This decay is correlated with the loss of regularity in the 
behavior of the atomic inversion that happens when neighboring 
revivals overlap. For larger numbers of atoms, the behavior of 
the atomic inversion is very irregular, and so is the behavior 
of squeezing. In particular, for $N=10$, the oscillations of 
$\xi$ decay already after $\tau_{R}/2$. 
The phenomenon of the collapses and revivals is quite regular 
for very large values of $\bar{n}$. We see in Fig.\ \ref{sq01}(b)
that for $\bar{n} = 900$ the squeezing parameter also behaves
very regularly. The structure of the oscillations of $\xi$
becomes more complicated (but keeps the regularity) as $N$
increases. For all values of $N$, the squeezing parameter
reaches deep minima at times just before integer multiples of 
$\tau_{R}/2$. The larger the value of $N$, the closer is the 
minimum of $\xi$ to $\tau_{R}/2$. (For $N \geq 6$, we also find
additional minima of $\xi$, but they are not so deep.)

The phenomenon of field squeezing for the half-excited Dicke 
state is drastically different from what we found for the 
fully-excited and ground states. In the former case ($m=0$), 
for moderate values of $\bar{n}$ ($\alpha < 30$), the minima 
of $\xi$ decrease below $1$ (i.e., squeezing occurs) only for 
$N=2$. This behavior is in contrast to the situation in the 
latter case ($m = \pm j$), where squeezing can be achieved for 
any $N$ (with a proper choice of $\alpha$) and is, moreover,
enhanced by increasing $N$. However, just two atoms prepared 
in the half-excited Dicke state can produce quite strong 
squeezing. 

\begin{figure}[htbp]
\epsfxsize=0.45\textwidth
\centerline{\epsffile{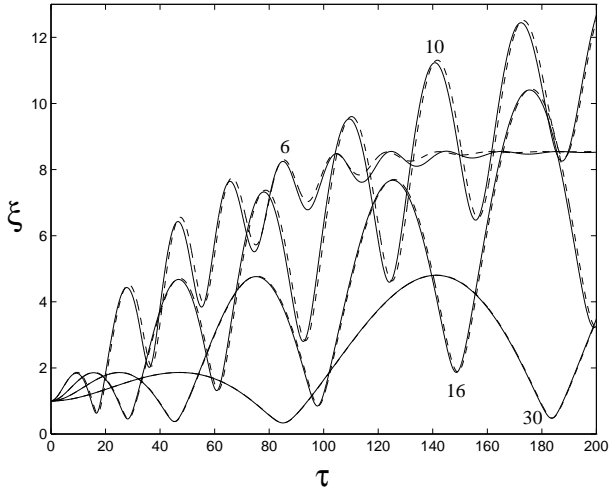}}
\vspace*{1mm}
\caption{The squeezing parameter $\xi$ versus the scaled time 
$\tau = g t$ for the initial coherent field state with
$\alpha = 6,10,16,30$ interacting with two atoms prepared in 
the half-excited Dicke state: numerical results (solid line)
and the factorization approximation (dashed line).}
\label{sq02}
\end{figure}

In Fig.\ \ref{sq02} we see the temporal behavior of $\xi$
for $N=2$ and $\alpha = 6,10,16,30$. The minima of $\xi$,
which occur at times before integer multiples of $\tau_{R}/2$,
become deeper (i.e., squeezing improves) as $\bar{n}$ 
increases. It is very important to note that here squeezing is
achieved at minima of the oscillations on the long-time 
scale. (Fast oscillations of $\xi$ on the short-time scale,
which produce squeezing for $m = \pm j$, are negligible
for the case  $m = 0$.) Consequently, for $m=0$ the duration 
of squeezing is essentially longer than for $m = \pm j$. 

Actually, this kind of squeezing on the long-time scale 
obtained for two atoms in the half-excited Dicke state has
the same physical origin as revival-time squeezing in the
Jaynes-Cummings model \cite{KuMa88,WoGB93}. This long-time
squeezing can be explained by means of the factorization
approximation for the semiclassical atomic states \cite{GB91}.
This approximation is valid for a strong initial coherent field 
and for times short compared with $\tau_{0} \sim \bar{n}$.

In a semiclassical treatment, one replaces the boson operators 
of the field in the interaction Hamiltonian with $c$-numbers. 
The eigenstates $|p\rangle_{\rm at}$ of this semiclassical 
Hamiltonian (called the semiclassical states) are just the 
eigenstates of the operator $J_{+} + J_{-} = 2 J_{x}$ with the 
eigenvalues $\lambda_{p} = N - 2 p$ ($p=0,1,\ldots,N$). 
An atomic state (and, in particular, a Dicke state) can be 
expanded in the basis of the semiclassical states. 
The factorization approximation means that if the atoms are 
prepared initially in a semiclassical state, then the total 
wave function of the system can be approximately written as a
product of its field and atomic parts \cite{GB91,ChKlSM94}:
\begin{equation}
|\Psi_{p}(t)\rangle \approx |A_{p}(t)\rangle_{\rm at} 
\otimes |\Phi_{p}(t)\rangle_{\rm f} .
\end{equation}

Using the factorization approximation, one can analytically
estimate mean values of the field operators corresponding
to a specific initial semiclassical atomic state 
$|p\rangle_{\rm at}$. In particular, one obtains
(as usual, we assume that $\alpha = \sqrt{\bar{n}}$ is real):
\begin{eqnarray}
\langle a \rangle_{p} & = & \alpha \exp\left[ - 
\frac{ i \lambda_{p} \tau }{ 2 \alpha } \left( 1- 
\frac{1}{4 \alpha^{2}} \right) \right] \nonumber \\
& & \times \exp\left\{ \alpha^2 \left[ 
e^{ i \lambda_{p} \tau/4 \alpha^{3} } - 1 - 
\frac{ i \lambda_{p} \tau }{ 4 \alpha^{3} } \right] \right\} ,
\label{ap}
\end{eqnarray}
\begin{eqnarray}
\langle a^{2} \rangle_{p} & = & \alpha^2 \exp\left[ - 
\frac{ i \lambda_{p} \tau }{ 2 \alpha } \left( 2- 
\frac{1}{\alpha^{2}} \right) \right] \nonumber \\
& & \times \exp\left\{ \alpha^2 \left[ 
e^{ i \lambda_{p} \tau/2 \alpha^{3} } - 1 - 
\frac{ i \lambda_{p} \tau }{ 2 \alpha^{3} } \right] \right\} .
\label{a2p}
\end{eqnarray}
Also, $\langle a^{\dagger} \rangle_{p}$ and 
$\langle a^{\dagger 2} \rangle_{p}$ are given by complex 
conjugates of $\langle a \rangle_{p}$ and 
$\langle a^{2} \rangle_{p}$, respectively, while 
$\langle a^{\dagger} a \rangle_{p} = \bar{n}$ can be taken
constant. Then an approximate expression for the squeezing
parameter can be easily obtained. For the semiclassical 
state $|p\rangle_{\rm at}$, one finds \cite{RSKC97}
\begin{eqnarray}
\xi_{p}^{2} & = & 1 + \left[ 2 \bar{n} \left( e^{-T_{p}^{2}/8}
- e^{-T_{p}^{2}/16} \right) \right. \nonumber \\
& & \left. + \frac{ T_{p}^{2} }{16} \left( e^{-T_{p}^{2}/16} - 
4 e^{-T_{p}^{2}/8} \right) \right] \cos\left( \frac{
\lambda_{p} \tau}{ \sqrt{\bar{n}} } \right) \nonumber \\ 
& & + \frac{T_{p} \sqrt{\bar{n}}}{2} \left( 2 e^{-T_{p}^{2}/8} 
- e^{-T_{p}^{2}/16} \right) \sin\left( \frac{\lambda_{p} \tau
}{ \sqrt{\bar{n}} } \right) \nonumber \\
& & + 2 \bar{n} \left( 1 - e^{-T_{p}^{2}/16} \right) ,
\label{xip-scl}
\end{eqnarray}
where $T_{p} = \lambda_{p} \tau / \bar{n}$. It can be seen that
squeezing is achieved for any semiclassical state, except
for those with $\lambda_{p} = 0$ (then $\xi_{p} = 1$ is 
constant as long as the factorization approximation is valid).
For $\lambda_{p} \neq 0$, the oscillations of the squeezing
parameter achieve minima for times
\begin{equation}
\tau_{\rm sq}^{(p)} \approx 0.9 \tau_{R}^{(p)}, 
1.95 \tau_{R}^{(p)},\ldots,
\end{equation}
where the revival time for the state $|p\rangle_{\rm at}$
is given by
\begin{equation}
\tau_{R}^{(p)} = \frac{ 2\pi \sqrt{\bar{n}} }{ |\lambda_{p}| }
\approx \frac{ \tau_{R} }{ |\lambda_{p}| } .
\end{equation}
The minima of $\xi_{p}$ become deeper (i.e., squeezing 
improves) as $\bar{n}$ increases.

Using formulas (\ref{ap}) and (\ref{a2p}), one can calculate
an approximate expression for the squeezing parameter for
an initial atomic state which is a superposition of the
semiclassical states. In particular, we will be interested
in approximate results for squeezing behavior of the Dicke 
states. First, note that in the Jaynes-Cummings model ($N=1$)
there are just two semiclassical states with 
$\lambda_{p} = \pm 1$, whose squeezing behavior is the same, 
for a good degree of accuracy. Consequently, any initial
atomic state in the Jaynes-Cummings model (and, in particular,
the ground and excited states) will exhibit the same squeezing 
behavior (within the validity of the factorization 
approximation). These considerations explain the appearance
of significant squeezing in the strong-field regime of the 
Jaynes-Cummings model for times near integer multiples of the 
revival time $\tau_{R}$. 

\vspace*{-3mm}
\begin{figure}[htbp]
\epsfxsize=0.45\textwidth
\centerline{\epsffile{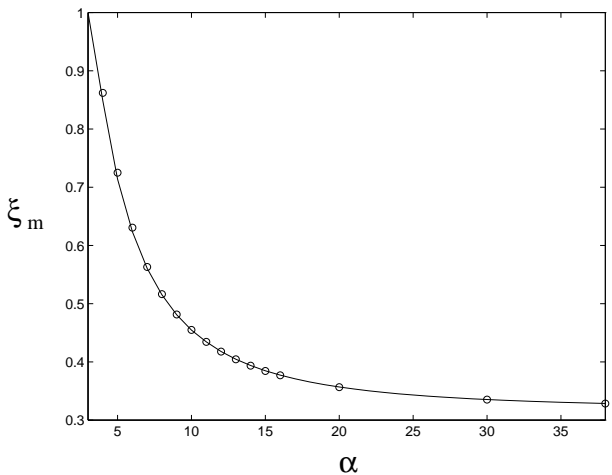}}
\vspace*{1mm}
\caption{The minimum value $\xi_{m}$ of the squeezing parameter 
versus the coherent amplitude $\alpha$ for two atoms prepared in 
the half-excited Dicke state: numerical results (circles) and 
the factorization approximation (line).}
\label{sq03}
\end{figure}

In the case of the Dicke model with two atoms, there are three 
semiclassical states; two of these states (with $\lambda_{p} = 
\pm 2$) lead to squeezing, while for the third state (with 
$\lambda_{p} = 0$) no squeezing is found ($\xi_{p} = 1$ is 
constant). The half-excited Dicke state $|1,0\rangle$ is a 
superposition of the two semiclassical states with $\lambda_{p} 
= \pm 2$, while the fully-excited and ground states also 
include the third semiclassical state with $\lambda_{p} = 0$. 
Consequently, the squeezing behavior of the half-excited Dicke 
state is very close to the behavior of a semiclassical state 
with $|\lambda_{p}| = 2$, while for the fully-excited and 
ground states squeezing is spoiled by the influence of the 
semiclassical state with $\lambda_{p} = 0$.
These considerations explain why squeezing on the long-time 
scale can be achieved only for the half-excited Dicke state
$|1,0\rangle$. The factorization approximation also explains
the appearance of the minima of the squeezing parameter at 
times close to integer multiples of $\tau_{R}/2$, whose depth 
increases with $\bar{n}$. From Fig.\ \ref{sq02} we see that
the factorization approximation describes very well the
squeezing behavior of the two-atom half-excited Dicke state
for large values of $\bar{n}$.
As shown in Fig.\ \ref{sq03}, the minimum value $\xi_{m}$ of 
the squeezing parameter decreases monotonically with $\alpha$. 
For $\bar{n}>10$, the factorization approximation is in 
excellent agreement with the numerical results. Similar
results are also obtained in the strong-field regime of the
Jaynes-Cummings model.
For $\alpha=38$ (the maximum value of $\alpha$ we 
considered), the best value of the squeezing parameter is 
$0.3285$. This is much better than the optimal values of 
squeezing which are obtained for the fully-excited or 
unexcited atoms when one takes $N \sim 10^{2}$. 

For atom numbers $N \geq 3$, any Dicke state will be a
superposition of the semiclassical states with different
values of $|\lambda_{p}|$. Since the minima of the squeezing
parameter $\xi_{p}$ occur at different times for different 
$|\lambda_{p}|$, squeezing on the long-time scale will be 
spoiled for the Dicke states of three and more atoms. It
can be shown \cite{RSKC97}, that for a Dicke state of 
$N$-atom system the condition for the existence of squeezing
in the revival regime is $\bar{n} > (2N)^{4}$. Of course,
the two-atom half-excited Dicke state is an exception to 
this rule, as it involves only two semiclassical states 
with the same value of $|\lambda_{p}|$. However, the above
condition is in good agreement with our numerical results
for $N \geq 3$. 

As we see, the factorization approximation can be very 
useful for explaining many features of the field-atom
interaction in the Dicke model. However, this approximation
fails to predict some interesting phenomena found by using
numerical calculations. In particular, squeezing on the 
short-time scale, which clearly dominates for the 
fully-excited and ground atomic states (especially, for 
large values of $N$), is not predicted by the factorization 
approximation. We also observe that the suppression of the
revival amplitude for the half-excited Dicke state cannot 
be described within this approximation. This can be 
readily understood by recalling the exact expression 
(\ref{popinv0}) describing the evolution of the population
inversion for the half-excited state $|1,0\rangle$. The
amplitude of the Rabi oscillations here is of the order 
of $1/\bar{n}$, which is neglected in the factorization 
approximation.

\section{Conclusions}

In this paper we considered in detail properties of the system
of $N$ two-level atoms interacting with a single-mode cavity
field (the Dicke model). When the field is initially in the
coherent state with a sufficiently large mean photon number, 
the dynamics of the system is quite regular, and the Rabi 
oscillations of the atomic inversion exhibit an interesting
quantum phenomenon of collapses and revivals.
We studied how this phenomenon is influenced by collective
atomic effects. The main conclusion is that the role of the
collective effects is determined by the initial atomic state.
We found that by preparing just two atoms in the half-excited 
Dicke state one can cause greater effect on the behavior of 
the system than by collecting tens or even hundreds of excited 
or unexcited atoms. In the phenomenon of the collapses and 
revivals, the half-excited Dicke state causes two basic effects: 
the revival amplitude is strongly suppressed (analogously
to the trapping phenomenon for a single atom in the equally
weighted superposition state) and the revival time is halved. 
The two-atom half-excited Dicke state also leads to very 
interesting squeezing behavior. It is the only Dicke state for 
$N \geq 2$ which exhibits strong squeezing on the long-time 
scale, similarly to the behavior found in the strong-field
regime of the Jaynes-Cummings model.

\acknowledgements

G.R. and C.B. gratefully acknowledge the financial help from 
the Technion. 
A.M. was supported by the Fund for Promotion of Research at 
the Technion and by the Technion VPR Fund.

\begin{table}
\caption{The ratio $f_{51} = h_{5}/h_{1}$, that represents 
the anharmonicity of the Hamiltonian eigenvalues, for
$n=36$, $N=10,20,30$ and $m=\pm j,0$.}
\label{T1}
\begin{tabular}{ccc}
$N$ & $m$ & $f_{51}$ \\ \hline
10 &      & 5.0219 \\
20 & $j$  & 5.0184 \\
30 &      & 5.0159 \\ \hline
10 &      & 5.0382 \\
20 & $-j$ & 5.0638 \\
30 &      & 5.1852 \\ \hline
10 &      & 5.0283 \\
20 & $0$  & 5.0309 \\
30 &      & 5.0360 
\end{tabular}
\end{table}

\end{document}